\documentstyle[12pt]{article}
\newcommand{\beq}{\begin{equation}}
\newcommand{\eeq}{\end{equation}}
\newcommand{\bea}{\begin{eqnarray}}
\newcommand{\eea}{\end{eqnarray}}

\renewcommand{\b}{\beta}

\newcommand{\n}{\nu}
\newcommand{\m}{\mu}

\newcommand{\s}{\sigma}

\newcommand{\dg}{\dagger}
\newcommand{\non}{\nonumber}

\newcommand{\rf}[1]{(\ref{#1})}

\begin{document}

\addtolength{\baselineskip}{0.20\baselineskip}

\hfill hep-lat/9504011

\hfill April 1995

\begin{center}

\vspace{24pt}

{ {\Large \bf Creutz Ratios From Color-Truncated Lattice
Configurations}}
\end{center}

\vspace{12pt}

\begin{center}

{\sl G. Fleming\footnote{email: gfleming@stars.sfsu.edu}
and J. Greensite\footnote{email: greensit@stars.sfsu.edu} }

\vspace{12pt}

Physics and Astronomy Dept. \\
San Francisco State University \\
1600 Holloway Ave. \\
San Francisco, CA 94132

\vspace{48pt}

{\bf Abstract}

\end{center}

\vspace{12pt}

   We investigate whether information about Creutz ratios is
encoded, separately, in each gluon color component of numerically
generated lattice configurations.  Working in SU(2) lattice gauge
theory in Landau gauge, we set two of the three gluon color components
to zero, and compensate for the loss of two-thirds of the fluctuation
by simply rescaling the remaining component by a factor of
$\sqrt{3}$.  Creutz ratios are then computed with this "abelianized"
configuration.  We find that the Creutz ratios of loops constructed
from abelianized links converge to the usual Creutz ratios in
the scaling regime.

\vfill

\newpage

    It has been shown in a number of studies \cite{Suzuki} that
if the link configurations of SU(2) lattice gauge theory are
transformed to the maximal abelian gauge (in which links are
maximally diagonal \cite{mag}), and then the link variables are
"abelianized" by truncation
to a set of diagonal components, that the Creutz
ratios computed from these "abelian" links match the Creutz
ratios computed from the full link variables quite closely.
The motivation of this procedure was to test the monopole
confinement mechanism proposed by 't Hooft \cite{tHooft} (for a
critical discussion, see ref. \cite{LMJ}).

   In this note we address the question of whether the reproduction
of Creutz ratios by color-truncated configurations is unique to the
maximal abelian gauge, or whether this is a property
of other gauges as well.  If the latter, it would suggest that information
about the QCD string tension is encoded, separately, in
{\it each} color component of the gluon field, at least in certain
gauges.\footnote{By a "color component" we are referring to gluon
color degrees of freedom in the adjoint representation, i.e. 3 components
for $SU(2)$, 8 components for $SU(3)$, etc.}

   We consider Landau gauge fixing in D=4 SU(2) lattice gauge theory;
the gauge-fixing condition is that
\beq
      \sum_{\mbox{\bf x}} \sum_{\m=1}^4 \mbox{Tr}[ U_\m(\mbox{\bf x}) +
U^{\dg}_\m (\mbox{\bf x}-\mbox{\bf e}_\m)]
              ~~~~~~\mbox{is maximized}
\eeq
with $SU(2)$ link variables,
\beq
      U_\m(x) = a_0 I + ia_k \s^k~~~~~~~\sum_{k=0}^3 a_k^2 = 1
\eeq
and $\s^k$ the Pauli matrices.
Having numerically generated and (Landau) gauge-fixed a lattice configuration,
by the standard Monte Carlo methods, we might try to "abelianize"
the configuration by setting
\beq
         a_1 = a_2 = 0
\label{trunc}
\eeq
and then rescaling $U$ to restore unitarity.  However, such a procedure
could hardly be expected to reproduce Creutz ratios, since it basically
just throws away most of the link fluctuation.  In maximal abelian gauge
this is not such a difficulty, because in that gauge it is normally
the case that
\beq
     a_3^2 >> a_1^2 + a_2^2
\eeq
i.e. most of the fluctuation is in the
color 3-direction.  In Landau gauge, however,
\beq
       <a_1^2> = <a_2^2> = <a_3^2>
\eeq
so the truncation \rf{trunc} is much more severe.  To mimic the effect
of the truncated degrees of freedom using the remaining degree of
freedom, let us just rescale $a_3$ by a constant
\beq
            a'_3 = \sqrt{3} a_3
\eeq
so that
\beq
         <(a'_3)^2> = <a_1^2> + <a_2^2> + <a_3^2>
\eeq
and then rescale each link variable to restore
unitarity.  The proposal is therefore to associate, with each
link variable $U$, an "abelianized" link variable $U'$ according
to the rule
\beq
      U = a_0 I + ia_k \s^k ~~~~ \Longrightarrow ~~~~
            U' = {a_0 I + i \sqrt{3} a_3 \s^3 \over a_0^2 + 3a_3^2}
\label{rule}
\eeq
and then calculate Wilson loops constructed from the abelianized
variables.

   With this prescription, we have calculated Creutz ratios $\chi[I,I]$
for loops built from the abelianized $U'$ links, up to $I=4$ and
$\beta=2.7$, on a $12^4$ lattice. The lattice was thermalized for 5000
iterations, data for loops up to $3 \times 3$ was taken every every
5th iteration of the next 5000 iterations; data for $\chi(4,4)$
was taken every 10th iteration of the next 40000 iterations following
thermalization.   The results, for the Creutz ratios
of the abelianized loops (open triangles) compared to the usual Creutz ratios
(solid triangles) are shown in Figure 1.  The convergence of the abelianized
ratios to the usual ratios, at larger $\b \ge 2.3$, is quite clear.

   Since there is nothing special about the 3-color in Landau gauge,
we conclude that somehow each gluon color in this gauge has encoded,
separately, information about the Creutz ratios.  We do not, at this
point, interpret this result as either supporting or discrediting
any particular confinement mechanism, although it suggests that the
maximal abelian gauge is not unique in reproducing Creutz ratios from
color truncated configurations.  Our data indicates only that, in Landau
gauge, the magnetic disorder which must account for an area law falloff
of Wilson loops appears to be present in each color separately, and
that the full disorder is obtained from the disorder of any single
component by a naive rescaling of that component by a factor
of $\sqrt{3}$.

    It is not hard to understand, for small loops at weak couplings,
why the $\sqrt{3}$ rescaling would reproduce the standard Wilson loop
values.  At weak couplings, in Landau gauge, Wilson loops
\beq
      W(C) = \mbox{Tr}<P \exp\left[i g \oint A^a_\m
                          {\s^a \over 2} dx^\m \right] >
\eeq
can be approximated by
\beq
      W(C) \approx \mbox{Tr}\exp \left[ - {g^2 \over 8} \oint dx^\m
                         \oint dy^\n <A^a_\m(x) A^a_\n(y)> I \right]
\label{approx}
\eeq
and in this approximation
\bea
      W(C) &=& \mbox{Tr}\exp \left[ - {g^2 \over 8} \oint dx^\m
                         \oint dy^\n \; 3 <A^3_\m(x) A^3_\n(y)> I \right]
\non \\
           &=& \mbox{Tr}\exp \left[ - {g^2 \over 8} \oint dx^\m
                         \oint dy^\n <A'^a_\m(x) A'^a_\n(y)> I \right]
\eea
where the abelianized field configuration
\bea
       A'^a_\m(x) = \left\{ \begin{array}{cl}
                           \sqrt{3} A^3_\m(x) & a=3 \\
                                0         &  a \ne 3 \end{array} \right.
\eea
corresponds, at weak couplings, to the color truncation rule in
eq. \rf{rule}.  Therefore, it would not be surprising to find that,
e.g., one-plaquette loops in the scaling region constructed from
abelianized links agree with the usual one-plaquette values.  The
agreement for small loops is simply a consequence of the fact that the
gluons are weakly interacting at small scales.  The surprise is that
the abelianized links, constructed with a naive $\sqrt{3}$ rescaling,
produce $\chi$-ratios which are fairly close to the usual values up
to the largest ($4 \times 4$) loops studied here.  At these loop sizes,
and, e.g., $\b = 2.4$, the effect of confinement
should be manifest, and the gluons are expected to be highly
interactive.  Even if magnetic disorder is encoded in a single gluon
color component, it is not obvious why the naive rescaling produces
$\chi$-ratios (and therefore string tensions) which are so nearly
correct.

     Possibly the approximation \rf{approx} may have
some validity outside perturbation theory (at least in Landau gauge).
This would explain why a simple $\sqrt{3}$ rescaling gives nearly
the right answer.  It remains, however, to explain why eq. \rf{approx},
which is reminiscent of one-gluon exchange, has any validity at all
in a strongly coupled regime.  All we can say, at this point, is that
our data is consistent with that approximation.

\vspace{33pt}

\noindent {\Large \bf Acknowledgements}{\vspace{11pt}}

    J.G. is happy to acknowledge the hospitality
of the Lawrence Berkeley Laboratory.  This work is supported in part
by  the U.S. Dept. of Energy, under Grant No. DE-FG03-92ER40711.

\bigskip
\bigskip
\bigskip
\bigskip

\noindent {\Large \bf Figure Caption}
\bigskip
\bigskip

\begin{description}
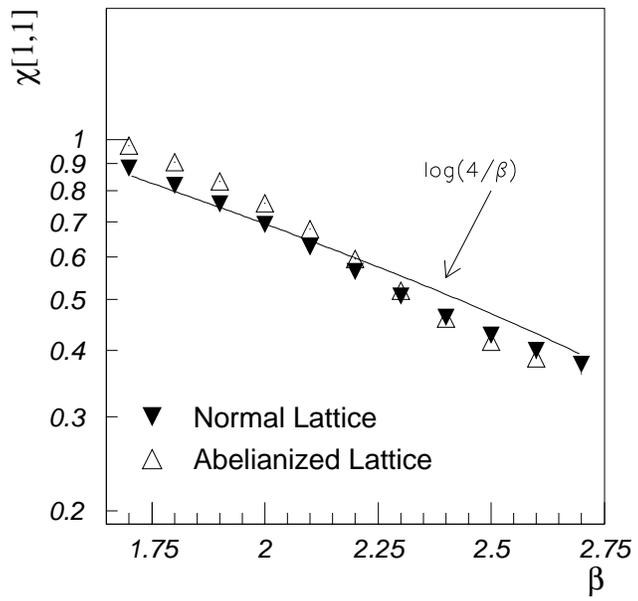
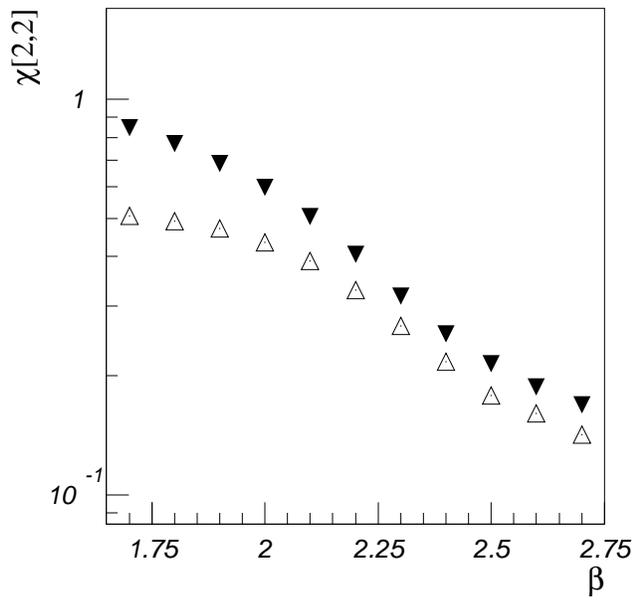
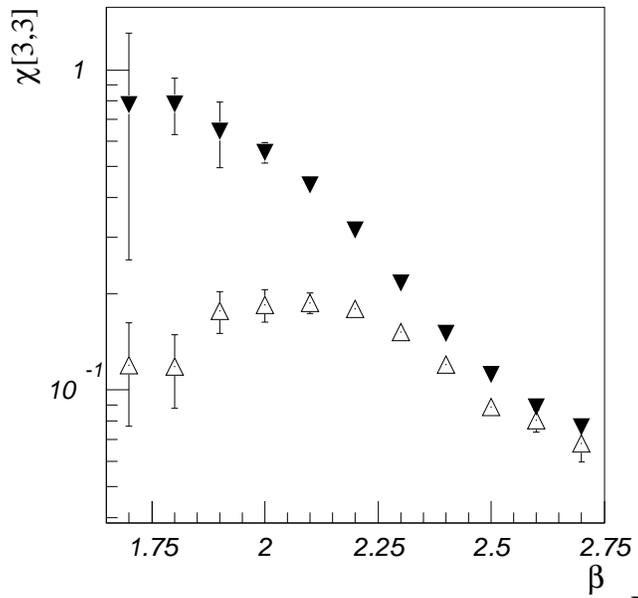
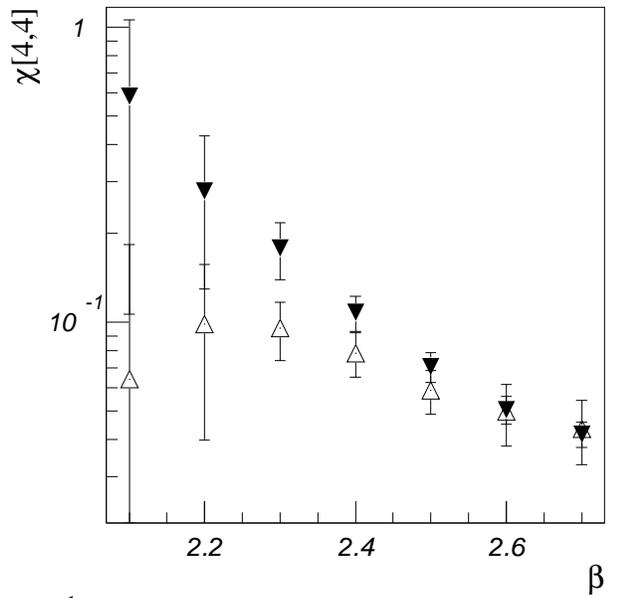
\item[Fig. 1] Creutz ratios $\chi[I,I]$ vs. $\b$, for $I=1-4$.
Solid triangles are
the usual ratios, open triangles are Creutz ratios obtained from
color-truncated ("abelianized") configurations.

\end{description}


\newpage

\begin{thebibliography}{xx}
\bibitem{Suzuki} T. Suzuki and I. Yotsuyanagi, Phys. Rev. D42 (1990) 4257, \\
                 S. Hioki et. al., Phys. Lett. B272 (1991) 326.
\bibitem{mag} A. Kronfeld, M. Laursen, G. Schierholz, and U.-J. Wiese,
              Nucl. Phys. B 293 (1987) 461.
\bibitem{tHooft} G. 't Hooft, Nucl. Phys. B138 (1978) 1.
\bibitem{LMJ} L. DelDebbio, M. Faber, and J. Greensite, Nucl. Phys. B414
              (1994) 594.
\end{thebibliography}
\end{document}